# Direct Role of Structural Dynamics in Electron-Lattice Coupling of Superconducting Cuprates

Fabrizio Carbone[1], Ding-Shyue Yang[1], Enrico Giannini[2], and

Ahmed H. Zewail[1*]

[1]Physical Biology Center for Ultrafast Science and Technology, Arthur Amos Noyes Laboratory of Chemical Physics, California Institute of Technology, Pasadena, CA 91125, USA.
[2] DPMC, University of Geneva, Quai E. Ansermet 24, 1211 Geneva CH.

**The mechanism of electron pairing in high-temperature superconductors is still the subject of intense debate. Here, we provide direct evidence of the role of structural dynamics, with selective atomic motions (buckling of copper-oxygen planes), in the anisotropic electron-lattice coupling. The transient structures were determined using time-resolved electron diffraction, following carrier excitation with polarized femtosecond heating pulses, and examined for different dopings and temperatures. The deformation amplitude reaches 0.5% of the *c*-axis value of 30 Å when the light polarization is in the direction of the copper-oxygen bond, but its decay slows down at 45º. These findings suggest a selective dynamical lattice involvement with the anisotropic electron-phonon coupling being on a time scale (1 to 3.5 ps depending on direction) of the same order of magnitude as that of the spin exchange of electron pairing in the high-temperature superconducting phase.**

\body

The pairing of electrons is now accepted as being essential in the formation of the superconducting condensate in high-temperature superconductors. What is debatable is the nature of forces ('glue') holding the pairs (1). The mechanism is different from that of conventional superconductors; for them, loss of the electric resistance is due to phonon-mediated electron pairing (Bardeen-Cooper-Schrieffer, BCS) (2). Ceramic cuprates become superconductors when extra holes or electrons are doped into their magnetically ordered charge-transfer insulator (ground) state (3, 4); the highest transition temperature ($T_c$) occurs at a doping of 0.15 extra hole per copper ion and it increases with the number (*n*) of Cu–O planes per unit cell, reaching a maximum at *n* = 3 (5). Because of the *d*-wave symmetry of the superconducting gap (6), the relatively small isotope effect (7, 8), and the magnitude of electron repulsion (*U*) and exchange (*J*) (appropriate for the antiferromagnetic phase), magnetic interactions have been considered as the source of binding (1, 9). On the other hand, the role of phonons in pairs formation has been discussed, from both experimental and theoretical perspectives (10, 11).

Angle-resolved photoemission spectroscopy (ARPES) experiments revealed the presence of kinks in the band dispersion at energies corresponding to specific (optical) phonon modes (7, 11–13). In some samples, inelastic neutron scattering data at similar energies supported a magnetic resonance mode below the transition temperature (14). The issue was raised over whether the low-energy features observed in the ARPES spectra are induced by magnetic or structural bosonic coupling. Based on energetics, the out-of-plane motion of the oxygen ions in the Cu–O plane, referred to as the out-of-plane buckling mode, has been assigned as responsible for the kink in the band dispersion observed along the direction of Cu–O bonds (11, 12). However, the electron-phonon coupling strength obtained by means of angle-integrated probes is not particularly large (15). This finding, together with the *d*-wave symmetry of the superconducting-phase order parameter (6), has been among the main arguments against a lattice-mediated pairing mechanism, because BCS theory of electron-phonon coupling favors an *s*-wave order parameter (16).

Theoretical calculations have suggested that selective optical-phonon excitation could lead to an anisotropic electron-phonon coupling (11). In cuprate superconductors, it has been demonstrated, using time-resolved photoemission (15) and time-resolved optical reflectivity (17), that the excited charge carriers preferentially couple to a phonon subset before decaying through anharmonic coupling to all other vibrations of the lattice. Moreover, the anomalous superconductivity-induced transfer of optical spectral weight and its doping dependence (17–19), usually considered as a hallmark of a non-BCS pairing mechanism, can be accounted for within a BCS model combined with a *d*-wave order parameter, for certain band structures (20). It is now known that the strength of the pairing potential (estimated from Nernst effect experiments) decreases as extra oxygen is doped into the unit cell, whereas the coherence length of the Cooper pairs increases with doping (21). The net effect is that for the condensate, a subtle compromise between pairing interaction and coherence needs to be achieved in order for high-temperature superconductivity to occur. To date, there has not been a direct observation of the actual structural change, or the anisotropy of electron-phonon coupling, in the superconducting state.

Here, we report, using time-resolved electron diffraction, the temporal evolution of the structure following polarized carrier excitation by a fs pulse, for different temperatures (for the metallic and superconducting states) and doping levels (from underdoped to optimally doped). Specifically, we investigated different compositions by varying the doping level and number of Cu–O planes per unit cell in the BSCCO (Bi, Sr, Ca, Cu, O) family; seven different crystals for a total of 30 cleavages were studied. The initial fs excitation transfers the system from the superconducting to the metallic state (15), breaking pairs (17). With the electron and lattice temperatures being vastly different (see below), energy of carriers is lowered through electron-phonon coupling which can be defined not only for the metallic but also for the superconducting state phase (22).

By varying the polarization of carrier excitation we observed major differences in the decay of the (00) diffraction rod which is correlated with the *c*-axis structural dynamics. The deduced structural changes on the time scale reported provide information on the mode(s) of atomic motions and the associated electron-phonon interactions. The striking polarization effect on the *c*-axis motion is consistent with a highly anisotropic electron-phonon coupling to the $B_{1g}$ out-of-plane buckling mode (50 meV), with the maximum amplitude of atomic motions being ~0.15 Å. The anisotropy follows the symmetry of the *d*-wave superconducting gap, with the largest coupling along the Cu–O bond where the gap has its maximum value. Along this direction the electron-phonon coupling parameter is obtained to be $\lambda = 0.55$, in the optimally doped two-layered sample, while at $45^o$ it is $\lambda = 0.08$. The previously reported value of 0.26 (15), within the framework of the Eliashberg formalism (23), thus represents an average over the different directions. More details of the experimental apparatus are given in the section of Methods and Experiments and in refs. 24 and 25.

**Results and Discussion.** We begin by discussing the results obtained for Bi2212. In Fig. 1, the static diffraction patterns of a single crystal of optimally doped Bi2212 are displayed. The patterns were recorded in the reflection geometry with the electron beam directed along three different axes, namely the [010], [110], and [100] directions, as displayed in panels A to C. The diffraction rods, which display the 2D nature of probing (Fig. 1, caption), were indexed for the tetragonal structure, giving the in-plane lattice parameters of $a = b = 5.40$ Å, consistent with X-ray values. The lattice modulation is resolved along the *b*-axis with a period of 27 Å, again in agreement with the X-ray data (26). The in-plane lattice constants, as well as the modulation, were confirmed for the specimens studied using our electron microscope, and one micrograph is given in Fig. 1*D*.

The temporal evolution of diffraction frames (with polarized excitation) is sensitive to motions of atoms during the structural change. In Fig. 2*A*, the intensity decay, due to motions of the ions (Debye-Waller effect), of the (00) rod is plotted for three different polarizations ($\vec{E}$) of the excitation pulse: $\vec{E}$ //[010], the direction of Cu–O bonds; $\vec{E}$ //[110], the direction at 45°; and the one at 22°. The data were taken at $T = 50$ K. At longer times, up to 1 ns, these transients recover very slowly; because of the poor *c*-axis conductivity and metallic *ab*-plane, heat transport is mainly lateral, but is complete on the time scale of our pulse repetition time (1 ms). In Fig. 2*B*, another set of data was obtained by rotating the sample while keeping the polarization parallel to the electron beam direction. The temporal evolution of the (00) diffraction intensity obtained from the two different orientations (electron beam parallel to the Cu–O bond, see diffraction pattern in Fig. 1*A*, and at 45°, Fig. 1*B*) shows the same anisotropic behavior as that obtained by rotating the polarization, ruling out possible experimental artefacts. We also verified that different diffraction orders show changes which scale with the scattering vector, confirming that the observed changes in the diffraction intensity originate from structural dynamics (see SI).

The intensity decay for different polarizations was found to have distinct time constants (see below): the decay is faster when the polarization is along the Cu–O bond and slows down when polarization is along the [110] direction (45° from the Cu–O bond). When charge carriers are excited impulsively through light in a crystal, the electron and lattice temperatures are driven out of equilibrium, but they equilibrate through electron-phonon coupling. Excitation of phonons causes the diffraction intensity to change with time, and this decrease mirrors an increase of the mean atomic displacement in the corresponding direction, with a temperature assigned to the displacement through a time-dependent Debye-Waller factor:

$$\ln[I(t)/I_0] = -2W(t) = -s^2 \langle \delta u^2(t) \rangle / 3, \qquad [1]$$

where $I(t)$ is the intensity of rod diffraction at a given time *t* after excitation, $I_0$ is the intensity before excitation, *s* is the scattering vector, and $\langle \delta u^2(t) \rangle$ is the mean square atomic displacement[†]. From the results reported here for $[I(t)/I_0]_{min}$, the root-mean-square value for the amplitude of the motion is obtained to be ~0.15 Å for 20 mJ/cm² fluence. Given the *c*-axis distance of 30 Å, this represents a change of 0.5%; the Cu–O planes separate by 3.2 Å.

The observed anisotropy of decay with polarization indicates the distinct *c*-axis distortion and the difference in electron-phonon coupling. In order to obtain the magnitude of the couplings we shall invoke the well-known model of electrons and lattice temperatures, dividing the lattice modes into those which are strongly coupled to the electrons and the rest which are not. Thus, the decrease of *I* at a given time tracks the change of $<\delta u^2(t)>$ with an effective corresponding temperature. For a Debye solid, the atomic displacement can be expressed as

$$\left\langle \delta u^2(t) \right\rangle = \frac{9\hbar^2 \Delta T(t)}{M k_B \Theta_D^2}, \qquad [2]$$

where *M* is the average mass in the unit cell, $k_B$ is Boltzmann constant, $\hbar$ is the reduced Planck constant, and $\Theta_D$ is the Debye temperature of the material (27). In Fig. 3*A*, we plot the equivalent temperature associated with the *c*-axis displacement of the optimally doped Bi2212 sample, for different polarizations, together with theoretical predictions given by the three-temperature model (15); see also SI. The initial heating of the charge carriers by the excitation pulse is on the fs time scale, during which an electronic temperature, $T_e$, is established. The coupling of carriers to a subset of phonon modes defines an equivalent temperature, $T_p$, for that phonon subset, and, subsequently, the relaxation to all other modes establishes the lattice temperature, $T_l$. It follows that the stronger the electron-phonon coupling the faster the decay of the diffraction intensity. From the results in Fig. 3*A*, we obtained $\lambda = 0.08$ for $\vec{E}$//[110] and 0.55 for $\vec{E}$//[010] in optimally doped Bi2212 (see SI). The average value at optimal doping is in good agreement with the results ($\lambda = 0.26$) of ref. 15, which angularly integrates the photoemission among different crystallographic directions, and also in agreement with "frozen-phonon" calculations (28).

The rate of diffraction change provides the time scales of selective electron-phonon coupling and the decay of initial modes involved. In Fig. 3*B*, the derivatives of the diffraction intensity as a function of time, d*I*(*t*)/d*t*, are displayed for different polarizations (Fig. 2*A*). The presence of a clear inversion point reflects the two processes involved, the one associated with the coupling between excited carriers and optical phonons, and the second that corresponds to the decay of optical modes, by anharmonicity into all other vibrations. The minimum in the derivative, signalling the crossover between these two processes, shifts toward an earlier time when the polarization becomes along the Cu–O bond. In Fig. 3*C*, the derivative of the simulated lattice temperature within the three-temperature model, d$T_l(t)$/d$t$, shows a similar two-process behavior. The clear shift of the minimum to an earlier time can be reproduced by varying the electron-phonon coupling parameter $\lambda$; in contrast, a change in the anharmonic coupling constant $\tau_a$ does not affect the early process, and the corresponding time of the derivative minimum has little shift (Fig. 3*C*, *inset*). Thus, consistent with the results of Fig. 3*A*, this analysis indicates that the anisotropic behavior of the diffraction intensity is indeed due to a directional electron-phonon coupling.

The derivative minima occur at times of ~1.0, 2.0 and 3.5 ps for, respectively, the polarization at 0°, 22° and 45° with respect to the Cu–O bond direction (Fig. 3*B*). Theoretically, the initial rate of the electron-phonon scattering can be obtained through the equation (23):

$$\frac{1}{\tau_{el-ph}} = \frac{3\hbar\lambda\langle\omega^2\rangle}{\pi k_B T_e}\left(1 - \frac{\hbar^2\langle\omega^4\rangle}{12\langle\omega^2\rangle(k_B T_e)(k_B T_l)} + \cdots\right) \approx \frac{3\hbar\lambda\langle\omega^2\rangle}{\pi k_B T_e}, \quad [3]$$

where $\tau_{el-ph}$ is the characteristic coupling time constant and $\omega$ is the angular frequency of the coupled modes. Given the values of $\lambda$ (0.55, 0.18 and 0.08 in Fig. 3*C*), we obtained $\tau_{el-ph}$ to be 290 fs, 900 fs and 2.0 ps with an initial $T_e$ = 6000 K and $T_l$ = 50 K. In ref. 15, $\tau_{el-ph}$ was reported to be 110 fs for $T_e$ ~ 600 K. Given the difference in fluence, hence $T_e$, the values of $\tau_{el-ph}$ obtained here (see Eq. **3**) are in reasonable agreement with the average value obtained in ref. 15. It should be emphasized that, within such time scale for the electron-phonon coupling, the lattice temperature $T_l$ remains below $T_c$; in Fig. 3*A*, the temperature crossover ($T_l > T_c$) occurs at 2 to 3 ps. We also note that at our fluence the photon doping has similar charge distribution to that of chemical doping (29).

The influence of polarization on the (00) diffraction rod (which gives the structural dynamics along the *c*-axis) reveals the unique interplay between the in-plane electronic properties and the out-of-plane distortion. Among the high-energy optical phonons that are efficiently coupled at early times, the in-plane breathing and out-of-plane buckling modes are favored (Fig. 4*A*) (7, 12) because of their high energy and involvement with carrier excitation at 1.55 eV. Our observation of a faster *c*-axis dynamics when the polarization is along the Cu–O bond implies a selective coupling between the excitation of charge carriers and specific high-momentum phonons. A plausible scheme is the stronger coupling between the antinodal ([010]) charge carriers and the out-of-plane buckling vibration of the oxygen ions in the Cu–O planes‡.

Further information was obtained by studying different compositions (dopings and number of layers) and temperatures. In Fig. 2*C* we display the results obtained for an underdoped Bi2212 ($T_c$ = 56 K), also at two temperatures. The anisotropy is evident at low temperature, giving the values of $\lambda$ = 0.12 for $\vec{E}$ //[110] and 1.0 for $\vec{E}$ //[010]. However, at higher temperature, the decay of both polarizations is similar and reaches the fastest profile recorded. This behavior is understood in view of the two types of phonons present at high temperature, those created through carrier-phonon coupling (low-temperature) and the ones by thermal excitation. This behavior with temperature is consistent with the optical reflection studies made by Gedik *et al.* (30).

On the other hand, for optimally doped Bi2223, we observed no significant anisotropy even in the low temperature regime (Fig. 2*D*). In fact, the intensity decay of the (00) rod for light polarized along [110] becomes essentially that of the [010] direction (see Fig. 2, *B* and *D*). The electron-phonon coupling in Bi2223 is thus similar for both directions ($\lambda$ = 0.40), signifying that the out-of-plane buckling motions are coupled more isotropically to the initial carrier excitation, likely due to the somewhat modified band structure (e.g., larger plasma frequency; see ref. 31) from that of Bi2212. This observation is consistent with the more isotropic superconducting properties of Bi2223 (32). The screening effect for the inner Cu–O layer by the outer ones in Bi2223 (33), and the less structural anisotropy between the in-plane and out-of-plane Cu–O distances (32), might also play a role in the disappearance of the anisotropic electron-phonon coupling.

In Fig. 4*B*, the doping dependence of $\lambda$ and the anisotropy observed for different polarizations, $\Delta\lambda = \lambda_{[010]} - \lambda_{[110]}$ (obtained from repeated experiments on different samples and cleavages), are displayed, together with the qualitative trend of the upper critical field (Nernst effect) and coherence length (21). The similarity in trend with upper critical field behavior, which can be related to the pair correlation strength, is suggestive of lattice involvement especially in this distinct phase region where the spin binding is decreasing. In view of an alternative explanation for the doping dependence of the critical field (34), our observation of an anisotropic coupling between the lattice and different light polarization may also be consistent with the idea of a dichotomy between nodal and antinodal carriers, with the latter forming a charge-density wave competing with superconductivity (35). Future experiments will be performed for completing the trends up to the overdoping regime for different superconductor transitions (36).

**Conclusion.** Observation of atomic motions and the directional electron-phonon coupling suggests that structural dynamics is an integral part of the description of the mechanism of high-temperature superconductivity. The anisotropic carrier-phonon coupling, reaching its maximum along the Cu–O bond, and the distortion of the Cu–O planes, suggest a direct role for structural dynamics and considerations (37, 38) beyond simple 2D models. Recent theoretical work has incorporated lattice phonons in the *t-J* model to account for the observed optical conductivity (39), whereas new band structure calculations suggested that large and directional electron-phonon coupling can favor spin ordering (40). It is worth noting that the reported time scale of electron-phonon coupling (at the photon/chemical doping level used) is of the same order of magnitude as that of spin exchange (40 fs) in the undoped phase. This implies that both the magnetic interactions and lattice structural changes should be taken into account for the microscopic description of the pair formation. Because it is now possible to examine the influence of these structural effects in the superconducting phase, by means of ultrafast electron crystallography, it is hoped that the reported results here can stimulate the development of theoretical models that explicitly incorporate the role of atomic motions in the mechanism of high-temperature superconductivity.

**Materials and Experiments.** The investigated specimens were the following: optimally doped and underdoped two-layered $Bi_2Sr_2CaCu_2O_{8+\delta}$ (Bi2212, $T_c = 91$ K and $T_c = 56$ K), and optimally doped three-layered $Bi_2Sr_2Ca_2Cu_3O_{10+\delta}$ (Bi2223, $T_c = 111$ K); see supporting information (SI) for details. All samples were cleaved in situ at low temperature, under a pressure on the order of $10^{-10}$ mbar in order to obtain a clean surface. An ultrashort (120 fs) laser pulse was used in order to induce a temperature jump in the sample, and the far-field diffraction of electron pulses was used to monitor the structural dynamics. The delay time between the pulses of carrier excitation and electron probing was varied while monitoring the change of Bragg diffraction intensities. The polarization of the excitation pulse was rotated with a half-wave plate and made parallel to the *a-b* plane of the sample. In all these experiments, the fluence used ranged from a few mJ/cm$^2$ and up to 20 mJ/cm$^2$.

**Footnotes.**

† The temporal evolution of $\sqrt{\langle \delta u^2(t) \rangle}$, deduced from *I*(*t*) according to Eq. **1**, may be fitted by considering different mechanisms. For the case of, e.g., nonequilibrium phase transition in ultrafast melting, the model of inertial dynamics (41) can be invoked with $\sqrt{\langle \delta u^2(t) \rangle}$ being related to the velocity of the motion, giving a Gaussian dependance on time. The model gives a velocity to be 0.025 Å/ps, far less than the root-mean-square velocity of 1.45 Å/ps at 50 K. The appropriate description for the nonequilibrium dynamics here should consider the energy transfer from the photoexcited carriers to the optical phonons, as described in the text (see also ref. 42).

‡ Unlike semiconductors where selectivity of *k*-vector in transient optical absorption is defined, in conventional metals the intraband transition (Drude region) involves near isotropic phonon-assisted transitions. In cuprates, nonconventional intraband transitions, even within the Drude region, may exist (43, 44). Moreover, on the ultrashort time scale, phonons are selectively coupled. However, at 1.55 eV cuprates has a more complex structure of bands and a contribution from the interband charge transfer will define a Cu–O polarization anisotropy. Given this complexity we cannot with certainty assign such a direction.

**Acknowledgement.** This work was supported by the National Science Foundation and the Air Force Office of Scientific Research in the Gordon and Betty Moore center for physical biology at Caltech. The authors acknowledge Dr. Nuh Gedik for his initial contribution to the project, Prof. Nai-Chang Yeh, Dr. Luca Perfetti, Prof. Peter Armitage, Dr. Hajo J. A. Molegraaf, and Erik van Heumen for stimulating discussions, and Helmuth Berger for providing the underdoped sample of Bi2212. We particularly wish to thank Profs. Dirk van der Marel, Petra Rudolf and Majed Chergui for critical reading of the manuscript and helpful suggestions.

**References:**

1. Anderson PW (2007) Is there glue in cuprate superconductors? *Science* 316:1705–1707.

2. Bardeen J, Cooper LN, Schrieffer JR (1957) Microscopic theory of superconductivity. *Phys Rev* 106:162–164.
3. Pickett WE (1989) Electronic structure of the high-temperature oxide superconductors. *Rev Mod Phys* 61:433–512.
4. Horsch P, Stephan W (1993) in *Electronic properties of high-temperature superconductors*, eds Kuzmany H, Mehring M, Fink J (Springer, Berlin).
5. Scott BA, Suard EY, Tsuei CC, Mitzi DB, McGuire TR (1994) Layer dependence of the superconducting transition temperature of $HgBa_2Ca_{n-1}Cu_nO_{2n+2+\delta}$. *Physica C* 230:239–245.
6. Van Harlingen DJ (1995) Phase-sensitive tests of the symmetry of the pairing state in the high-temperature superconductors–Evidence for $d_{x^2-y^2}$ symmetry. *Rev Mod Phys* 67:515–535.
7. Gweon GH *et al.* (2004) An unusual isotope effect in a high-transition-temperature superconductor. *Nature* 430:187–190.
8. Zech D *et al.* (1994) Site-selective oxygen isotope effect in optimally doped $YBa_2Cu_3O_{6+x}$. *Nature* 371:681–683.
9. Anderson PW (1987) The resonating valence bond state in $La_2CuO_4$ and superconductivity. *Science* 235:1196–1198.
10. Slezak JA *et al.* (2008) Imaging the impact on cuprate superconductivity of varying the interatomic distances within individual crystal unit cells. *Proc Natl Acad Sci USA* 105:3203–3208.
11. Devereaux TP, Cuk T, Shen ZX, Nagaosa N (2004) Anisotropic electron-phonon interaction in the cuprates. *Phys Rev Lett* 93:117004.
12. Cuk T *et al.* (2004) Coupling of the $B_{1g}$ Phonon to the antinodal electronic states of $Bi_2Sr_2Ca_{0.92}Y_{0.08}Cu_2O_{8+\delta}$. *Phys Rev Lett* 93:117003.
13. Meevasana W *et al.* (2006) Doping dependence of the coupling of electrons to bosonic modes in the single-layer high-temperature $Bi_2Sr_2CuO_6$ superconductor. *Phys Rev Lett* 96:157003.
14. Hayden SM, Mook HA, Dai P, Perring TG, Dohan F (2004) The structure of the high-energy spin excitations in a high-transition temperature superconductor. *Nature* 429:531–534.
15. Perfetti L *et al.* (2007) Ultrafast electron relaxation in superconducting $Bi_2Sr_2CaCu_2O_{8+\delta}$ by time-resolved photoelectron spectroscopy. *Phys Rev Lett* 99:197001.
16. Annett J, Goldenfeld N, Renn SR (1991) Interpretation of the temperature dependence of the electromagnetic penetration depth in $YBa_2Cu_3O_{7-\delta}$. *Phys Rev B* 43:2778–2782.
17. Gedik N *et al.* (2005) Abrupt transition in quasiparticle dynamics at optimal doping in a cuprate superconductor system. *Phys Rev Lett* 95:117005.
18. Molegraaf HJA, Presura C, van der Marel D, Kes PH, Li M (2002) Superconductivity-induced transfer of in-plane spectral weight in $Bi_2Sr_2CaCu_2O_{8+\delta}$. *Science* 295:2239–2241.
19. Carbone F *et al.* (2006) Doping dependence of the redistribution of optical spectral weight in $Bi_2Sr_2CaCu_2O_{8+\delta}$ *Phys Rev B* 74:064510.
20. Marsiglio F, Carbone F, Kuzmenko AB, van der Marel D (2006) Intraband optical spectral weight in the presence of a van Hove singularity: Application to $Bi_2Sr_2CaCu_2O_{8+\delta}$. *Phys Rev B* 74:174516.
21. Wang Y *et al.* (2003) Dependence of upper critical field and pairing strength on doping in cuprates. *Science* 299:86–89.
22. Scalapino DJ, Schrieffer JR, Wilkins JW (1966) Strong-coupling superconductivity. I. *Phys Rev* 148:263–279.

23. Allen PB (1987) Theory of thermal relaxation of electrons in metals. *Phys Rev Lett* 59:1460–1463.
24. Zewail AH (2006) 4D ultrafast electron diffraction, crystallography, and microscopy. *Annu Rev Phys Chem* 57:65–103.
25. Baum P, Zewail AH (2006) Breaking resolution limits in ultrafast electron diffraction and microscopy. *Proc Natl Acad Sci USA* 103:16105–16110.
26. Giannini E, Garnier V, Gladyshevskii R, Flukiger R (2004) Growth and characterization of $Bi_2Sr_2Ca_2Cu_3O_{10}$ and $(Bi,Pb)_2Sr_2Ca_2Cu_3O_{10-\delta}$ single crystals. *Supercond Sci Technol* 17:220–226.
27. Stampfli P, Bennemann KH (1992) Dynamical theory of the laser-induced lattice instability of silicon. *Phys Rev B* 46:10686–10692.
28. Savrasov SY, Andersen OK (1996) Linear-response calculation of the electron-phonon coupling in doped $CaCuO_2$. *Phys Rev Lett* 77:4430–4433.
29. Gedik N, Yang D-S, Logvenov G, Bozovic I, Zewail AH (2008) Nonequilibrium phase transitions in cuprates observed by ultrafast electron crystallography. *Science* 316:425–429.
30. Gedik N *et al.* (2004) Single-quasiparticle stability and quasiparticle-pair decay in $YBa_2Cu_3O_{6.5}$. *Phys Rev B* 70:014504.
31. Carbone F *et al.* (2006) In-plane optical spectral weight transfer in optimally doped $Bi_2Sr_2Ca_2Cu_3O_{10}$. *Phys. Rev. B* 74:024502.
32. Giannini E *et al.* (2008) Growth, structure and physical properties of single crystals of pure and Pb-doped Bi-based high $T_c$ superconductors. *Curr Appl Phys* 8:115–119.
33. Kotegawa H *et al.* (2001) Unusual magnetic and superconducting characteristics in multilayered high-$T_c$ cuprates: $^{63}$Cu NMR study. *Phys Rev B* 64:064515.
34. Beyer AD, Chen C-T, Grinolds MS, Teague ML, Yeh N-C (2008) Competing orders and the doping and momentum dependent quasiparticle excitations in cuprate superconductors. *Physica C* 468:471–479.
35. Chen C-T, Beyer AD, Yeh N-C (2007) Effects of competing orders and quantum phase fluctuations on the low-energy excitations and pseudogap phenomena of cuprate superconductors. *Solid State Commun* 143:447–452.
36. Edwards PP, Mott NF, Alexandrov AS (1998) The insulator-superconductor transformation in cuprates. *J Supercond* 11:151–154.
37. Phillips JC (2008) Quantum percolation in cuprate high-temperature superconductors. *Proc Natl Acad Sci USA* 105:9917–9919.
38. Tahir-Kheli J, Goddard III, WA (2007) Chiral plaquette polaron theory of cuprate superconductivity. *Phys Rev B* 76:014514.
39. Mishchenko AS, *et al.* (2008) Charge dynamics of doped holes in high-$T_c$ cuprate superconductors: A clue from optical conductivity. *Phys Rev Lett* 100:166401.
40. Jarlborg T (2003) Spin-phonon interaction and band effects in the high-$T_c$ superconductor $HgBa_2CuO_4$. *Phys Rev B* 68:172501.
41. Lindenberg AM *et al.* (2005) Atomic-scale visualization of inertial dynamics. *Science* 308:392–395.
42. Yang D-S, Gedik N, Zewail AH (2007) Ultrafast electron crystallography. 1. Nonequilibrium dynamics of nanometer-scale structures. *J Phys Chem C* 111:4889–4919.
43. Stephan W, Horsch P (1990) Optical properties of one- and two-dimensional Hubbard and *t-J* models. *Phys. Rev. B* 42:8736–8739.

44. Kuiper P *et al.* (1998) Resonant X-Ray Raman Spectra of Cu *dd* Excitations in $Sr_2CuO_2Cl_2$. *Phys. Rev. Lett.* 80:5204–5207, and references therein.

**Figure Legends:**

**Fig. 1.** Static diffraction patterns of optimally doped Bi2212. (*A–C*) Reflection patterns obtained at three different electron probing directions $\vec{v}_e$ (by rotating the crystalline sample), as indicated in the lower right corner. The large lattice constant along *c* and the nm-depth of electron probing give rise to the rod-like patterns; from (*A*), the intensity modulation along the diffraction rods gives the out-of-plane lattice parameter of *c* =30 Å. The indices for different diffraction rods are given. Note that the satellites of the main diffraction rods in (*C*) manifest the 27-Å modulation along the *b*-axis of Bi2212. (*D*) Transmission diffraction pattern obtained by our electron microscope. The square in-plane structure is evident, with the presence of the *b*-axis modulation which is also seen in (*C*).

**Fig. 2.** Time-resolved diffraction. (*A*) Diffraction intensity change of the (00) rod at different polarizations in optimally doped Bi2212. The laser fluence was 20 mJ/cm$^2$ and the temperature was 50 K. The electron probing was kept along [110] (Fig. 1*B*), and *θ* is the angle of polarization away from the probing direction (controlled by rotation of a half-wave plate). The dotted lines (and also those in panels *B* to *D*) show the fits to an apparent exponential decay. (*B*) Diffraction intensity change of the (00) rod, from the same sample, obtained with the optical polarization being parallel to the electron probing. By rotating the crystal, the time-dependent change was measured for the two zone axes (Figs. 1*A* and 1*B*). (*C*) Diffraction intensity change of the (00) rod for an underdoped Bi2212 sample ($T_c$ = 56 K), at two temperatures and two polarizations. The inset shows the diffraction pattern obtained from the sample, revealing the good quality of the crystal. (*D*) Diffraction intensity change obtained from a three-layered, optimally doped Bi2223 sample at 45 K for two polarizations. The diffraction pattern is displayed as an inset.

**Fig. 3.** Experimental and theoretical intensity transients. (*A*) Lattice temperature derived from diffraction using Eqs. **1** and **2**, for different polarizations, along [010] (blue dots) and [110] (red dots), in optimally doped Bi2212. From the three-temperature model described in the text and SI, we obtain $\lambda$ = 0.08 for $\vec{E}$//[110] ($T_l$, red solid line) and $\lambda$=0.55 for $\vec{E}$//[010] ($T_l$, blue solid line).

The electronic ($T_e$, dashed lines) and hot-phonon ($T_p$, solid lines) temperatures are also displayed. (*B*) Derivatives of the (00) diffraction intensity derived from Fig. 2*A* for different polarizations. (*C*) Derivatives of the simulated lattice temperature within the three-temperature model, for different $\lambda$ with a fixed anharmonic coupling time $\tau_a = 2.8$ ps [also shown in (*B*)] and for different $\tau_a$ with a fixed $\lambda = 0.26$ (*inset*). The clear shift of the minimum position is only observed when $\lambda$ is varied (black dotted lines).

**Fig. 4.** Crystal structure and the phase diagram. (*A*) Three-dimensional structure of Bi2212 (26), indicating the main crystallographic directions. Relevant to our work are the red arrows which show the atomic movements in the in-plane breathing mode (left panel) and those in the out-of-plane buckling mode (right panel). (*B*) The doping dependence of the coupling constant ($\lambda$) along the [010] and [110] directions in Bi2212 (blue and red dots, respectively) and its anisotropy ($\Delta\lambda$ between the two directions; black solid line), as well as $\lambda$ along the [010] and [110] directions in Bi2223 (green and orange dots, respectively) and the extrapolated anisotropy (black dashed line). A qualitative sketch of the upper critical field and Cooper-pair coherence length (green and violet lines) is also shown (21).

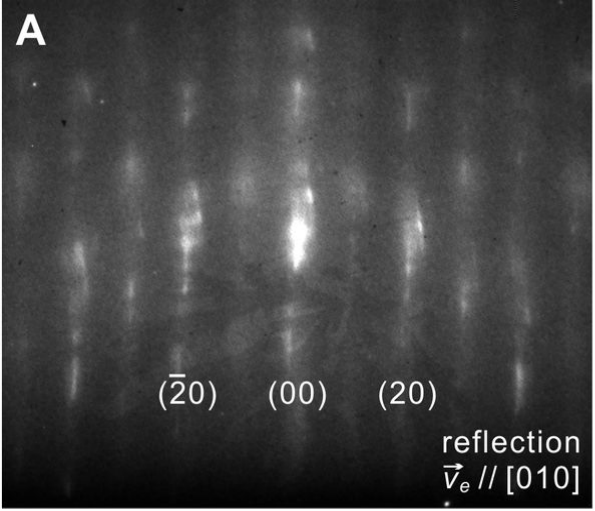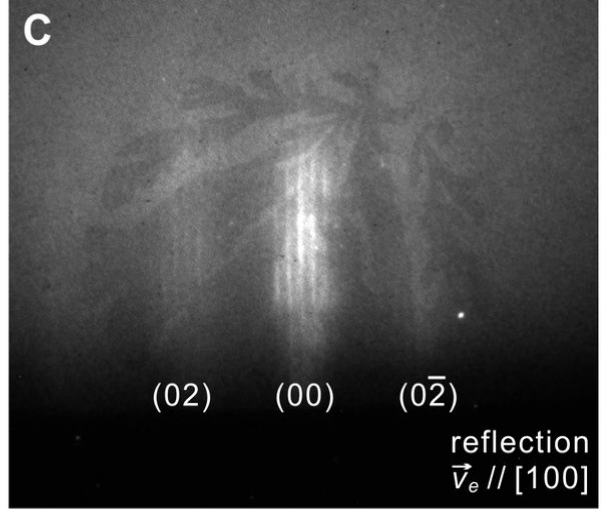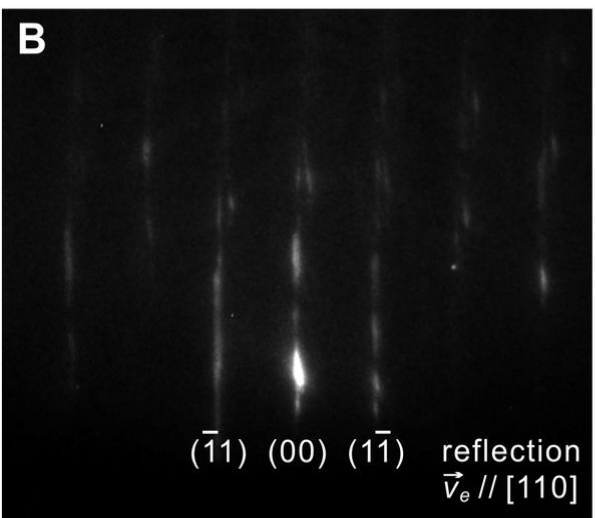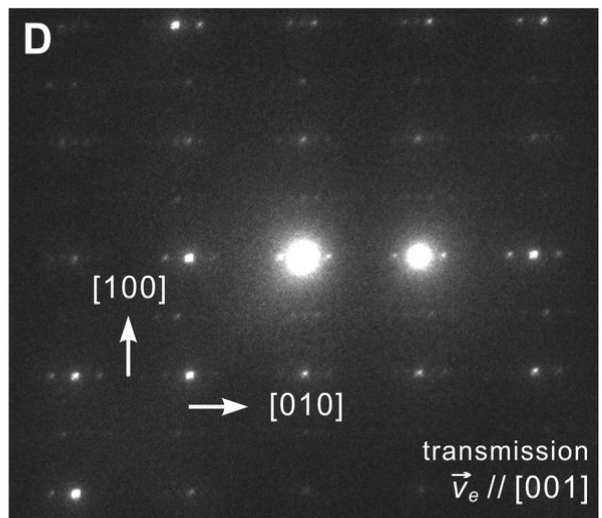

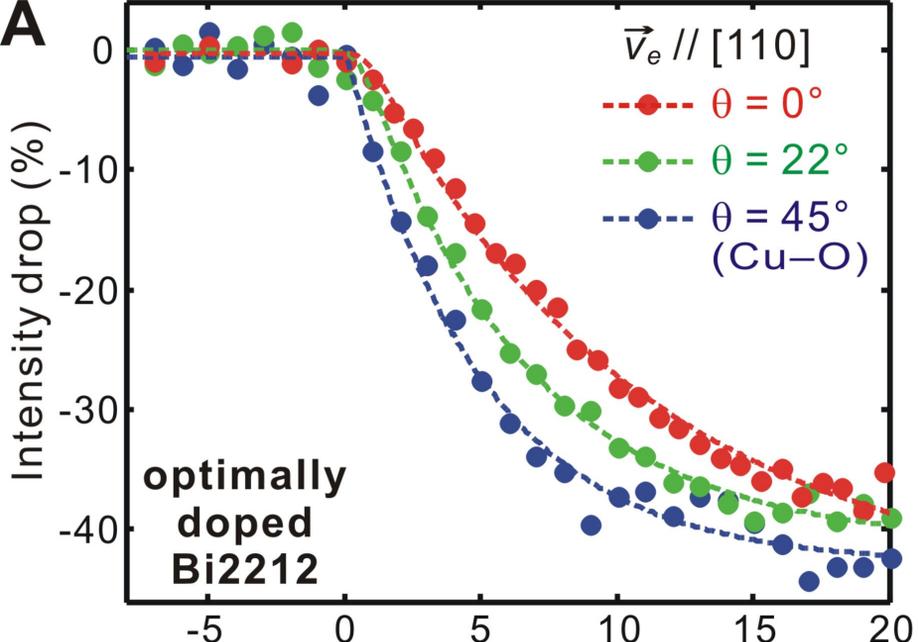
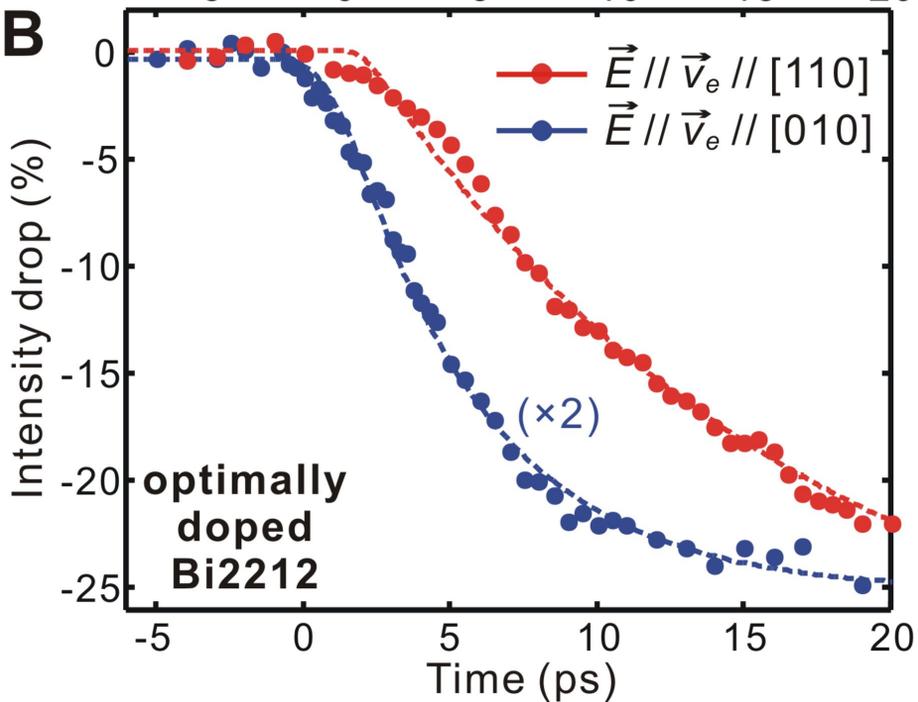
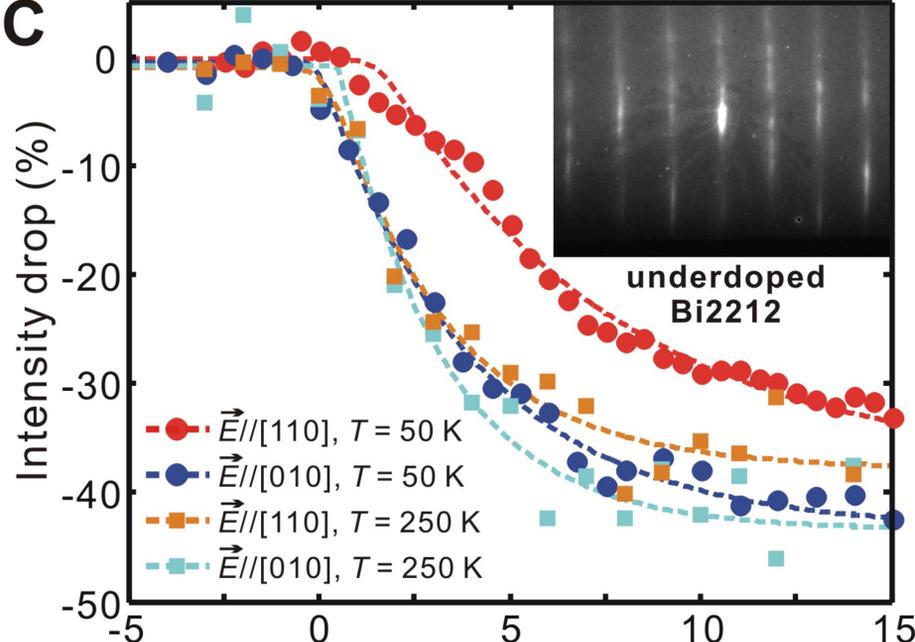
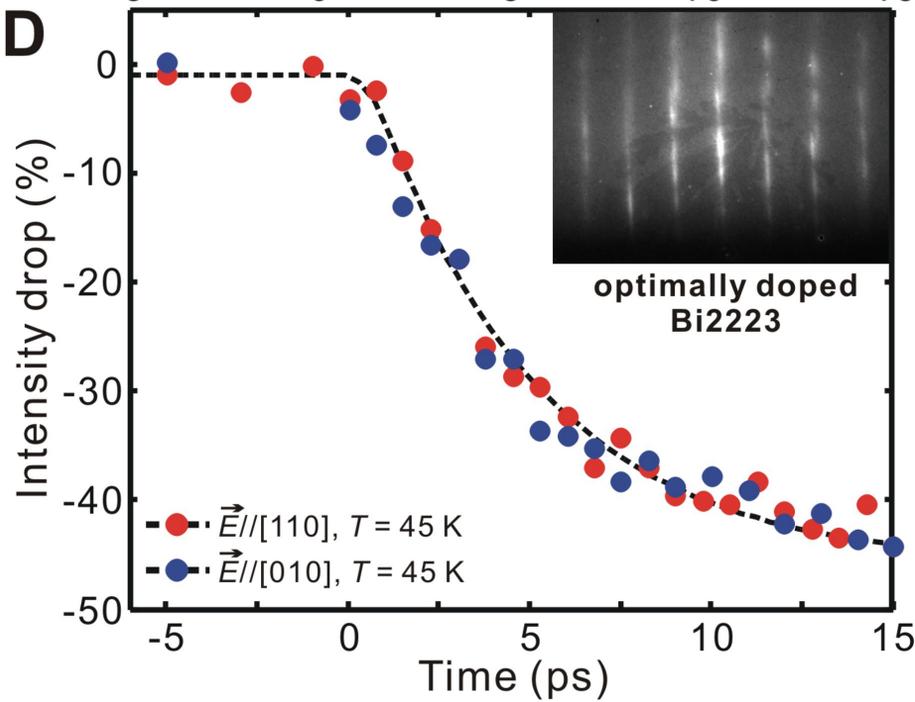

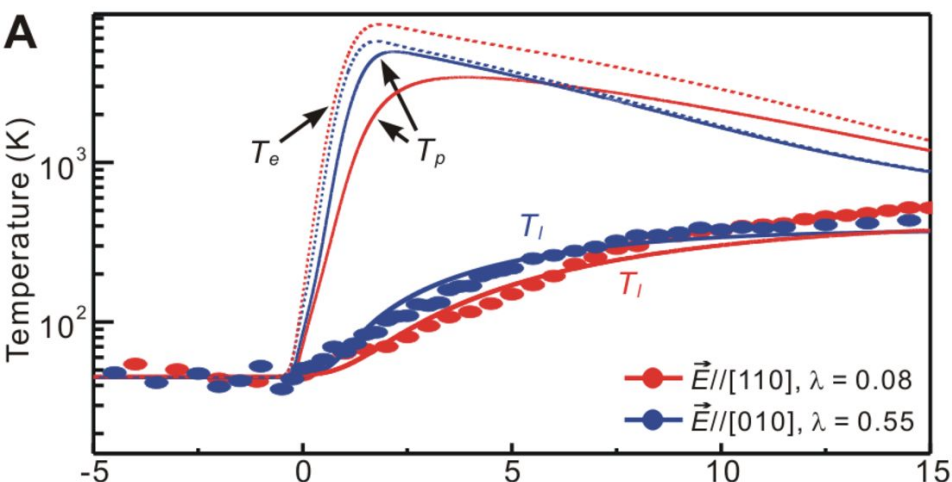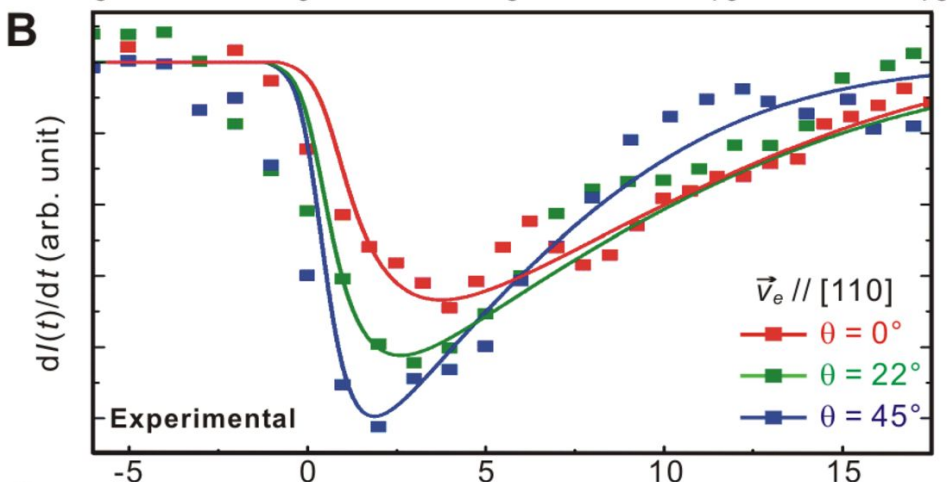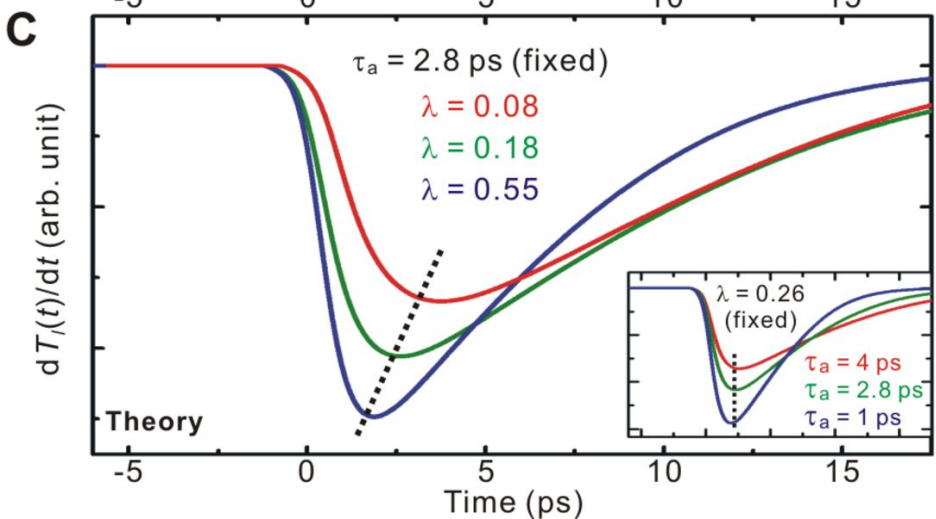

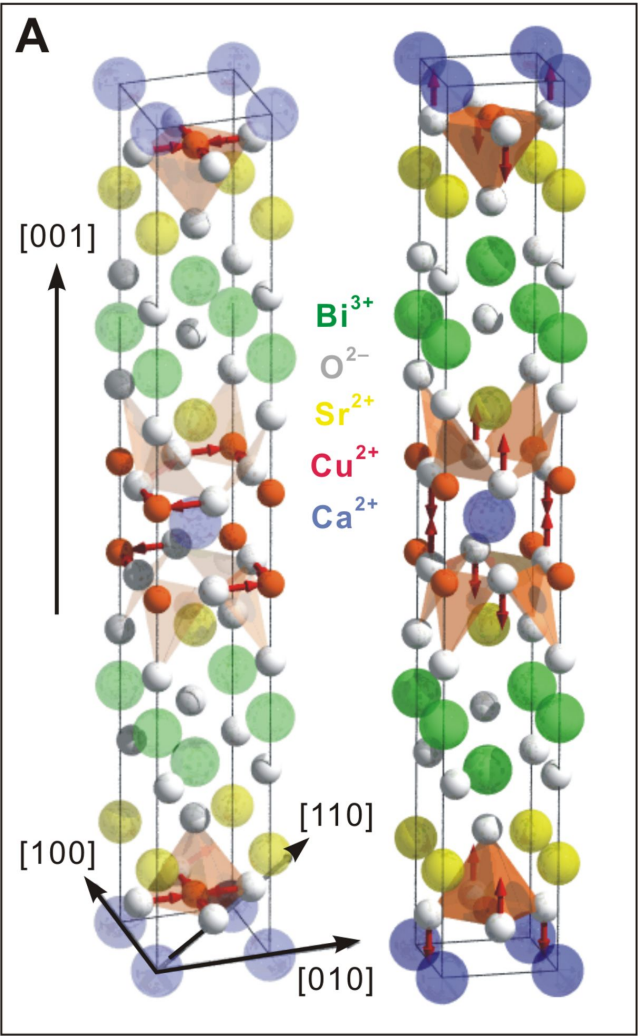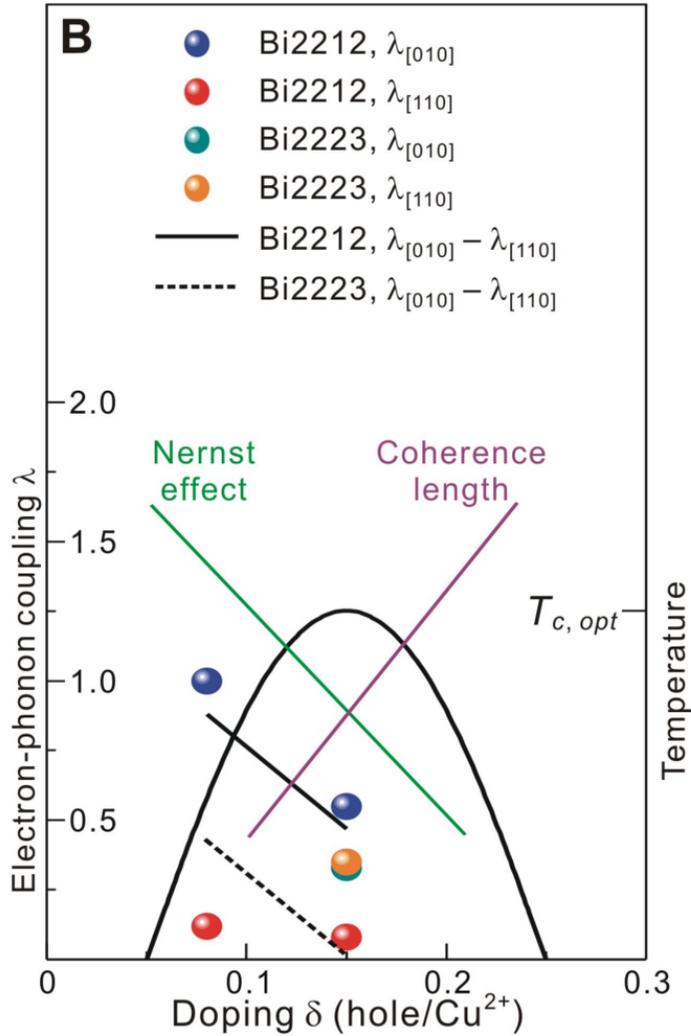

# Supporting Information

## Direct Role of Structural Dynamics in Electron-Lattice Coupling of Superconducting Cuprates

Fabrizio Carbone[1], Ding-Shyue Yang[1], Enrico Giannini[2], Ahmed H. Zewail[1]*

[1] *Physical Biology Center for Ultrafast Science and Technology, Arthur Amos Noyes Laboratory of Chemical Physics, California Institute of Technology, Pasadena, CA 91125, USA.*
[2] *DPMC, University of Geneva, Quai E. Ansermet 24, 1211 Geneva CH.*

### Materials and characterization

The optimally doped samples of Bi2212 and Bi2223 were grown by the travel solvent floating zone technique, described in ref. S1. The superconducting transition temperature was found to be $T_c$ = 91 K in Bi2212 ($\Delta T_c$ = 1 K), and $T_c$ = 111 K in Bi2223 ($\Delta T_c$ = 4 K). The underdoped Bi2212 sample was grown by the self-flux method (S2), annealed in an oxygen-deficient atmosphere, and its transition temperature was $T_c$ = 56 K ($\Delta T_c$ < 6 K). The magnetic susceptibility curves for two representative samples are displayed in Fig. S1.

### Intensity scaling of Bragg diffractions

In a time-resolved diffraction experiment, different Bragg spots at a given time should exhibit intensity changes in accord with the scattering vectors, $s$ (see Eq. **1**). Therefore, two distinct Bragg diffraction features appearing at $s = s_1$ and $s_2$ should obey the following scaling relation:

$$\frac{\ln(I_{s_1}/I_0)}{\ln(I_{s_2}/I_0)} = \left(\frac{s_1}{s_2}\right)^2. \tag{S1}$$

In Fig. S2, we display the intensity changes for two different Bragg spots, recorded in the same pattern, at different scattering vectors. The correct scaling relationship confirms that the observed intensity changes are indeed originated from structural motions.

### Three-temperature model

Conventionally, the two-temperature model (S3) is used to describe laser-induced heating of the electron and phonon subsystems in an elementary metal. Its success is the result of the isotropic electron-phonon coupling in a simple lattice structure, i.e., one atom per primitive unit cell. In complex, strongly correlated materials like high-$T_c$ superconductors, however, such model becomes inappropriate because photoexcited carriers may anisotropically and preferentially couple to certain optical phonon modes, resulting in the failure of assignment of a single temperature to the whole lattice structure.

In the three-temperature model described in ref. S4, in addition to the electron temperature $T_e$, two temperatures are defined for the lattice part: the hot-phonon temperature, $T_p$, for the subset of phonon modes to which the laser-excited conduction-band carriers transfer their excess energy, and the lattice temperature, $T_l$, for the rest of the phonon modes which are thermalized through anharmonic coupling. As an approximation, the spectrum of the hot phonons $F(\Omega)$ is assumed to follow an Einstein model: $F(\Omega) = \delta(\Omega-\Omega_0)$, where $\delta$ denotes the Dirac delta function, $\Omega$ the energy, and $\Omega_0$ the energy of a hot phonon. Effectiveness of the energy transfer between the carriers and hot phonons is described by the dimensionless parameter $\lambda$: $\lambda = 2\int \Omega^{-1}\alpha^2 F d\Omega$, where $\alpha^2 F$ is the Eliashberg coupling function (S3). The rate equations describing the temporal evolution of the three temperatures are given by:

$$\frac{dT_e}{dt} = -\frac{3\lambda\Omega_0^3}{\hbar\pi k_B^2}\frac{n_e - n_p}{T_e} + \frac{P}{C_e}, \tag{S1}$$

$$\frac{dT_p}{dt} = \frac{C_e}{C_p}\frac{3\lambda\Omega_0^3}{\hbar\pi k_B^2}\frac{n_e - n_p}{T_e} - \frac{T_p - T_l}{\tau_a}, \tag{S2}$$

$$\frac{dT_l}{dt} = \frac{C_p}{C_l}\frac{T_p - T_l}{\tau_a}, \tag{S3}$$

where $\tau_a = 2.8$ ps is the characteristic time for the anharmonic coupling of the hot phonons to the lattice, $n_e$ and $n_p$ are the electron and hot-phonon distributions given by $n_{e,p} = (e^{\Omega_0/k_B T_{e,p}} - 1)^{-1}$, and $P$ is the laser fluence function; a ratio of $10^3$ between the electronic specific heat $C_e$ and the lattice specific heat ($C_p$ and $C_l$) is known (S4).

In our calculations, the values of the parameters were chosen to be the same as in ref. S4, except for the excitation source which in our case has a fluence of 20 mJ/cm$^2$ and duration of 120 fs. The fit of the simulated lattice temperature to our data gives the following results for the electron-phonon coupling constant in the different samples: $\lambda_{[110]} = 0.12$, $\lambda_{[010]} = 1.0$ and their average $\lambda_{avg} = 0.56$ in underdoped Bi2212; $\lambda_{[110]} = 0.08$, $\lambda_{[010]} = 0.55$ and their average $\lambda_{avg} = 0.31$ in optimally doped Bi2212; $\lambda_{[110]} \approx \lambda_{[010]} = 0.40$ in optimally doped Bi2223. We provide the fits to the data for optimally doped Bi2212 and Bi2223 in Fig. S3.

**References:**

S1. Giannini E, Garnier V, Gladyshevskii R, Flukiger R (2004) Growth and characterization of $Bi_2Sr_2Ca_2Cu_3O_{10}$ and $(Bi,Pb)_2Sr_2Ca_2Cu_3O_{10-\delta}$ single crystals. *Supercond Sci Technol* 17:220–226.

S2. Nakamura N, Shimotomai M (1991) Growth of $YBa_2Cu_3O_x$ single crystals by a self-flux method with alkali chlorides as additives. *Physica C* 185:439–440.

S3. Allen PB (1987) Theory of thermal relaxation of electrons in metals. *Phys Rev Lett* 59:1460–1463.

S4. Perfetti L *et al.* (2007) Ultrafast electron relaxation in superconducting $Bi_2Sr_2CaCu_2O_{8+\delta}$ by time-resolved photoelectron spectroscopy. *Phys Rev Lett* 99:197001.

**Figure Legends**

**Fig. S1** Sample characterization. (A) Temperature dependence of the magnetic susceptibility of an optimally doped Bi2223 sample. (B) Temperature dependence of the magnetic susceptibility of a representative optimally doped Bi2212 sample investigated. The sharp transition attests the crystallinity and composition homogeneity of our samples.

**Fig. S2** Intensity scaling between two diffractions. Shown are the decay of two distinct Bragg peaks, observed at $s_1 = 6.3$ Å$^{-1}$ (red) and $s_2 = 4.5$ Å$^{-1}$ (blue). The green curve is obtained by multiplying the data at $s = s_2$ by the factor of $(s_1/s_2)^2$, according to Eq. (S1), and its match with the data at $s = s_1$ confirms the structurally induced diffraction changes following the carrier excitation.

**Fig. S3** Lattice temperatures derived from the diffraction intensity data using Eq. (1–2) for different polarizations, along [010] (blue dots) and [110] (red dots), in the underdoped Bi2212 (A) and in the optimally doped Bi2223 (B). Dashed and solid lines show the calculated temporal evolution of the three temperatures, $T_e$, $T_p$ and $T_l$.

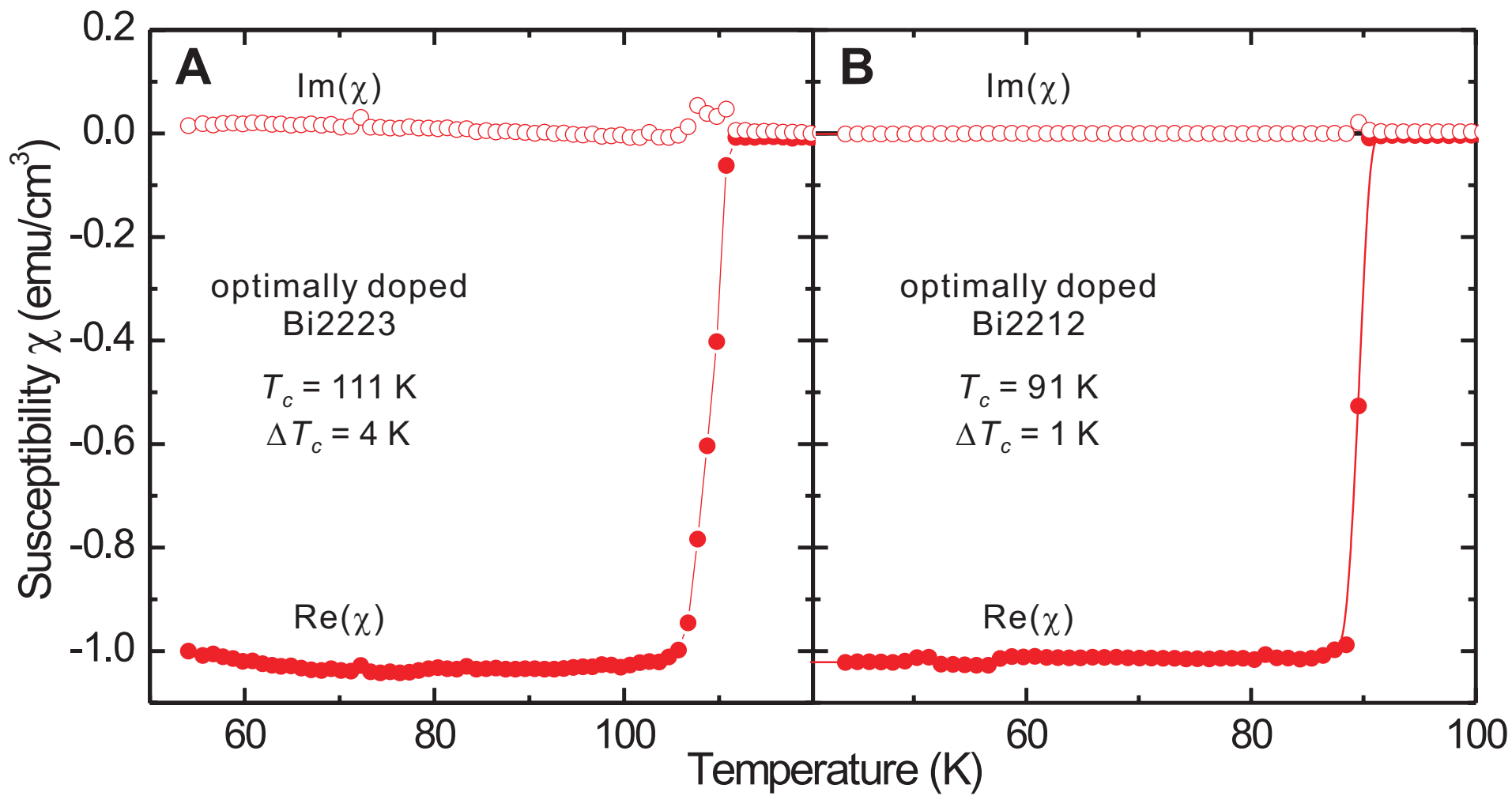

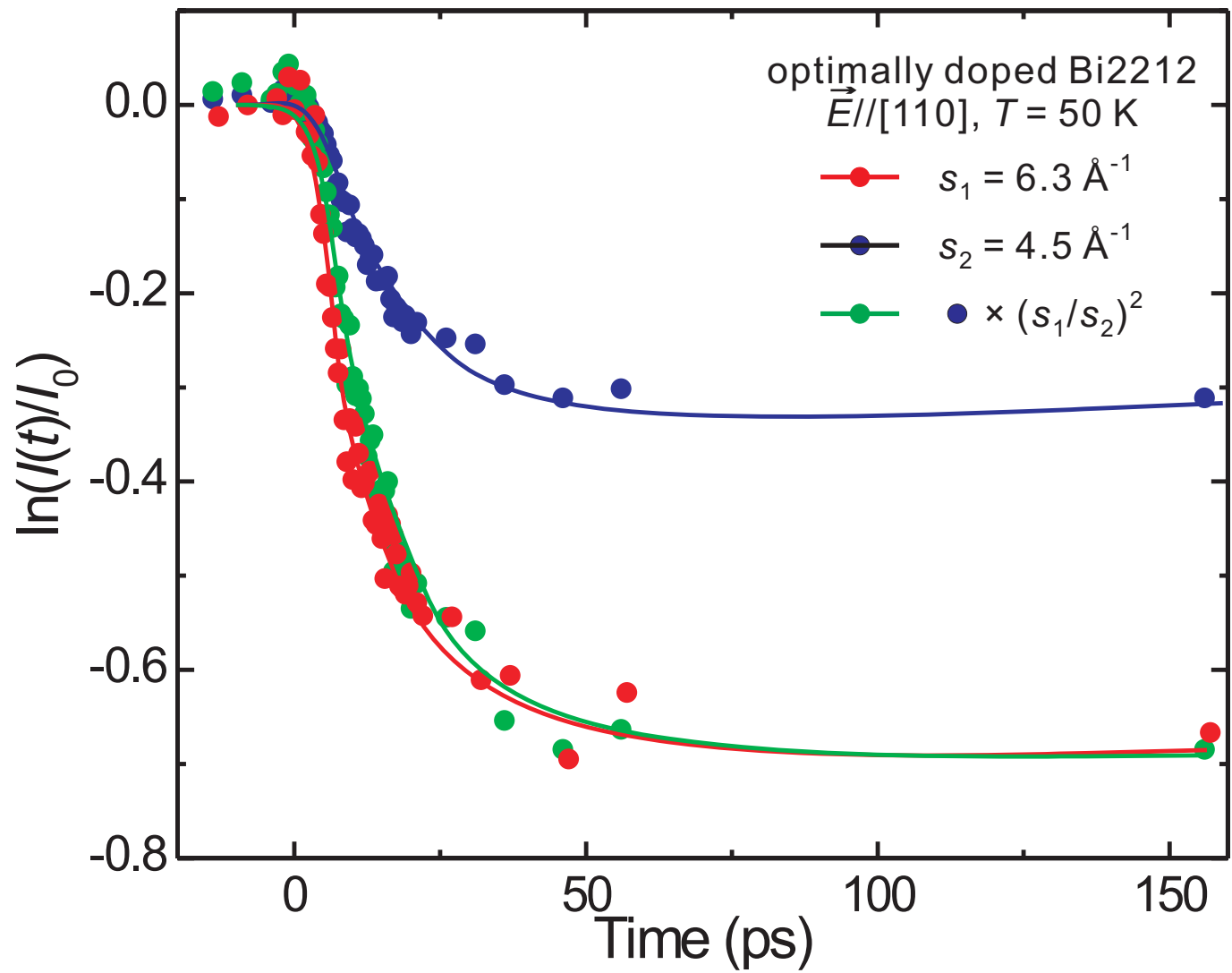

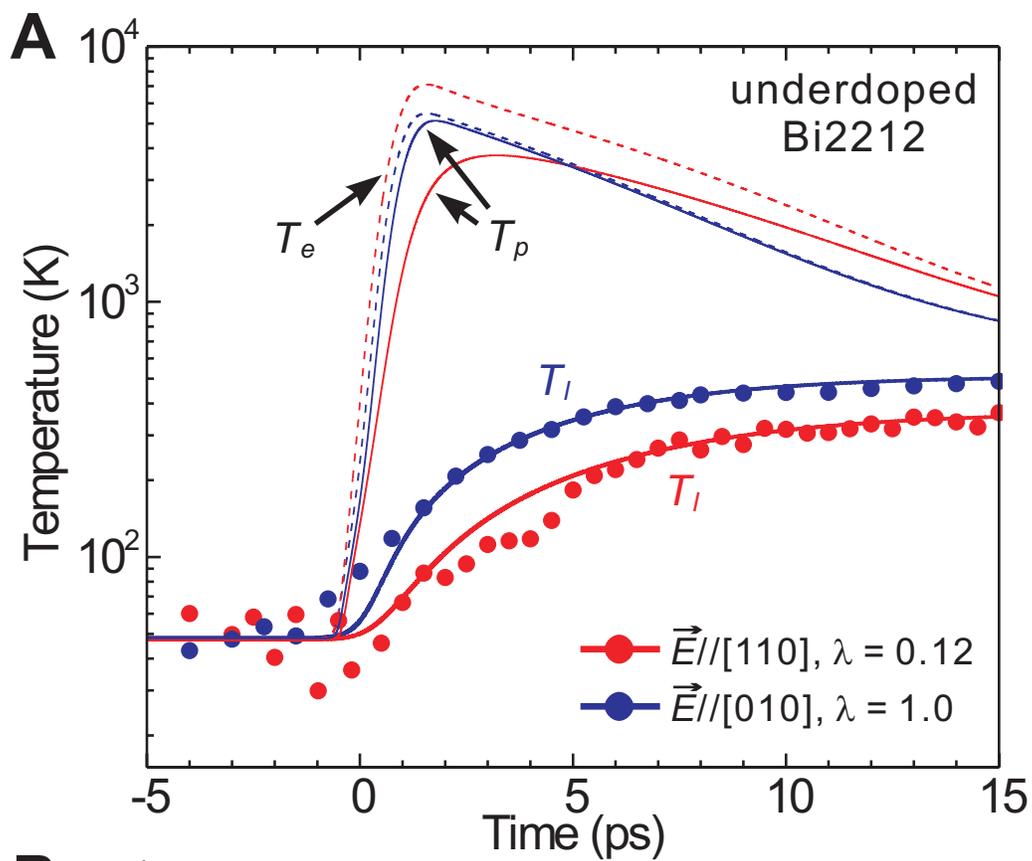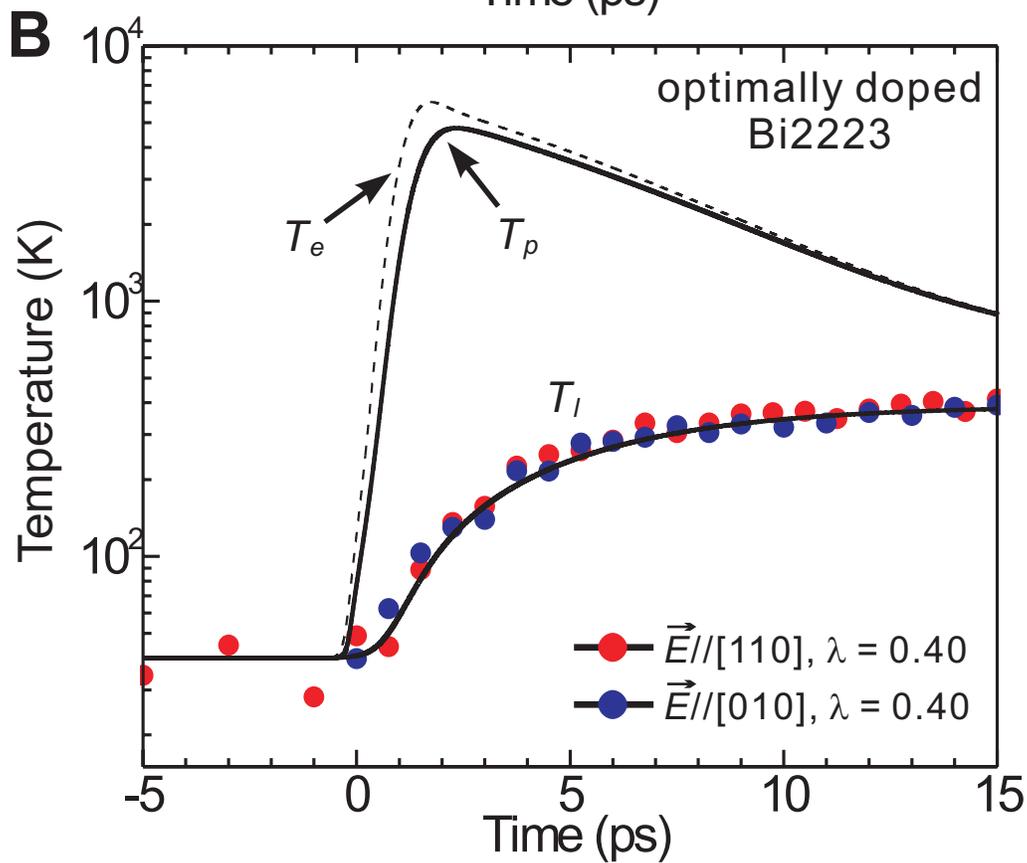